\documentclass[a4paper]{article}
\usepackage{graphicx}
\usepackage{amsmath}
\usepackage{amssymb} 

\begin{document}
\begin{titlepage}

\rightline{{\large \tt May 2002}}

\vskip 1.4 cm

\centerline{\Large \bf
Ordinary atom-mirror atom bound states: }
\centerline{\Large \bf
A new window on the mirror world 
}

\vskip 1.2 cm

\centerline{\large R. Foot$^a$ and S. Mitra$^b$}
\vskip 0.7 cm\noindent
\centerline{{\large $^a$\ \it foot@physics.unimelb.edu.au}}

\centerline{{\large \it School of Physics}}

\centerline{{\large \it Research Centre for High Energy Physics}}

\centerline{{\large \it The University of Melbourne}}

\centerline{{\large \it Victoria
3010 Australia }}
\vskip 0.3cm
\centerline{{\large $^b$\ \it saibalm@science.uva.nl}}
\centerline{{\large \it Instituut voor Theoretische Fysica}}
\centerline{{\large \it Universiteit van Amsterdam}}
\centerline{{\large \it 1018 XE Amsterdam}}
\centerline{{\large \it The Netherlands}} 

\vskip 1.4cm

\centerline{\large Abstract} \vskip 0.5cm \noindent
Mirror symmetry is a plausible candidate for a fundamental symmetry
of particle interactions which 
can be exactly conserved if a set of
mirror particles exist. The properties of the mirror 
particles seem to provide an excellent candidate to explain
the inferred dark matter of the Universe and might
also be responsible for a variety of other puzzles in 
particle physics, astrophysics, meteoritics and planetary science.
One such puzzle -- the orthopositronium lifetime problem --
can be explained if there is a small kinetic mixing of ordinary
and mirror photons. We show that this kinetic mixing implies the
existence of ordinary atom - mirror atom bound states with interesting
terrestrial and astrophysical implications.
We suggest that sensitive mass spectroscopic studies of ordinary
samples containing heavy elements such as lead might reveal the
presence of these bound states, as they would appear as anomalously
heavy elements.
In addition to the effects of single mirror atoms, collective
effects from embedded fragments of mirror matter 
(such as mirror iron microparticles) are also possible.
We speculate that such mirror matter fragments might
explain a mysterious UV photon burst observed coming from a laser
irradiated lead target in a recent experiment.

\end{titlepage}
%\vskip 0.1cm

One of the most natural candidates for a symmetry of nature is
parity symmetry (also called left-right or mirror symmetry). While
it is an established experimental fact that parity symmetry
appears broken by the interactions of the known elementary
particles, this however does not exclude the possible existence of
exact unbroken parity symmetry in nature. This is because parity
(and also time reversal) can be exactly conserved if a set of
mirror particles exist\cite{ly,flv}. The idea is that for each
ordinary particle, such as the photon, electron, proton and
neutron, there is a corresponding mirror particle, of exactly the
same mass as the ordinary particle
\footnote{ The mirror particles only have the same mass as 
their ordinary counterparts provided that the mirror symmetry is unbroken.
It is possible to write down gauge models where
the mirror symmetry is broken\cite{broken,ug}, in some cases
allowing the mirror particles to have completely
arbitrary masses\cite{ug}, however these scenarios
tend to be more complicated and much less well motivated in our view.}.
Furthermore, the mirror particles interact with each other in
exactly the same way that the ordinary particles do. It follows that
the mirror proton
is stable for the same reason that the ordinary proton is stable,
and that is, the interactions of the mirror particles conserve a
mirror baryon number. The mirror particles are not produced
(significantly) in laboratory experiments just because they couple
very weakly to the ordinary particles. In the modern language of
gauge theories, the mirror particles are all singlets under the
standard $G \equiv SU(3)\otimes SU(2)_L \otimes U(1)_Y$ gauge
interactions. Instead the mirror fermions interact with a set of
mirror gauge particles, so that the gauge symmetry of the theory
is doubled, i.e. $G \otimes G$ (the ordinary particles are, of
course, singlets under the mirror gauge symmetry)\cite{flv}.
Parity is conserved because the mirror fermions experience $V+A$
(right-handed) mirror weak interactions and the ordinary fermions
experience the usual $V-A$ (left-handed) weak interactions.
Ordinary and mirror particles interact with each other
predominately by gravity only.

At the present time there is a fascinating range of experimental
observations supporting the existence of mirror matter, for a
review see Ref.\cite{puz} (for a more detailed discussion
of the case for mirror matter, accessible to
the non-specialist, see the recent book\cite{bk}). 
The evidence includes numerous observations
suggesting the existence of invisible `dark matter' in galaxies.
Mirror matter is stable and dark and provides a natural candidate
for this inferred dark matter\cite{blin}. 
The MACHO observations\cite{mo}, close-in extrasolar
planets\cite{ce}, isolated planets\cite{is} and
even gamma ray bursts\cite{grb} may
all be mirror world manifestations.
In our solar system, the anomalous slowing down of the
Pioneer spacecraft\cite{pioneer}, the Tunguska event and some anomalous 
low altitude
fireball phenomena\cite{tunguska,yoon} have also been identified as possible 
manifestations of the mirror world.
On the quantum level, small fundamental
interactions connecting ordinary and mirror matter are possible.
Theoretical constraints from gauge invariance, renormalizability
and mirror symmetry suggest only three possible types of
interactions\cite{flv,flv2}: photon-mirror photon kinetic mixing,
neutrino-mirror neutrino mass mixing and Higgs-mirror Higgs
interactions. The main experimental implication of photon-mirror
photon kinetic mixing is that it modifies the properties of
orthopositronium, leading to a shorter effective lifetime in
`vacuum' experiments\cite{gl,gn,fg}. A shorter lifetime is in fact
seen at the 5 sigma level!\cite{vac,fg}. Neutrino-mirror neutrino
mass mixing implies maximal oscillations for each ordinary
neutrino with its mirror partner -- a result which
may be connected with the neutrino physics anomalies\cite{flv2,mm}. 
%This ma a simple and predictive explanation for the apparent $\sim 50\%$ solar
%$\nu_e$ flux reduction obtained in the solar neutrino experiments
%(the solar neutrino problem), as well as the observed
%$\approx 50\%$ reduction in upgoing $\nu_\mu$ in atmospheric 
%neutrino experiments.
% [Although it is also true that the mirror world solution to the
% neutrino physics anomalies is not
% in perfect agreement with all of the experimental neutrino data at
% the moment].

Of the few possible ways in which ordinary
and mirror matter can interact with each other besides
gravity, 
of most importance for this paper is
the photon-mirror photon kinetic mixing interaction.
In field theory,
photon-mirror photon kinetic mixing
is described by the interaction
\begin{equation}
{\cal L} = {\epsilon \over 2}F^{\mu \nu} F'_{\mu \nu},
\label{ek}
\end{equation}
where $F^{\mu \nu}$ ($F'_{\mu \nu}$) is the field strength tensor
for electromagnetism (mirror electromagnetism). This type of
Lagrangian term is gauge invariant and renormalizable and can
exist at tree level\cite{flv,fh} or may be induced radiatively in
models without $U(1)$ gauge symmetries (such as grand unified
theories)\cite{gl,bob,cf}. One effect of ordinary photon-mirror
photon kinetic mixing is to give the mirror charged particles a
small electric charge\cite{flv,gl,bob}. That is, they couple to
ordinary photons with electric charge $\epsilon e$.

The most important experimental implication of photon-mirror
photon kinetic mixing is that it modifies the properties
of orthopositronium\cite{gl}. This effect arises due to
radiative off-diagonal contributions to the
orthopositronium, mirror orthopositronium
mass matrix. This means that orthopositronium oscillates
into its mirror partner. Decays of mirror
orthopositronium are not detected
experimentally which effectively increases the observed
decay rate\cite{gl}. Because collisions of orthopositronium destroy
the quantum coherence, this mirror world effect is most
important for experiments which are designed such that the
collision rate of the orthopositronium is low\cite{gn}.
The only accurate experiment sensitive to the mirror
world effect is the Ann Arbour vacuum cavity experiment\cite{vac}.
This experiment obtained a decay rate of
$\Gamma_{oPs} = 7.0482 \pm 0.0016 \ \mu s^{-1}$.
Normalizing this measured value with the current
theoretical value of $7.0399 \ \mu s^{-1}$ \cite{theory} gives
\begin{equation}
{\Gamma_{oPs}(exp) \over \Gamma_{oPs}(theory)} =
1.0012 \pm 0.00023
\end{equation}
which is a five sigma discrepancy with theory.
It suggests a value $|\epsilon| \simeq 10^{-6}$
for the photon-mirror photon kinetic mixing\cite{fg}.
Taken at face value this experiment is strong evidence
for the existence of mirror matter and hence
parity symmetry. 
The sign of $\epsilon$ is not constrained by the orthopositronium
experiments and so there are two distinct possibilities depending
on whether $\epsilon > 0$ or $\epsilon < 0$.

Perhaps one of the most fascinating implications of the photon
- mirror photon kinetic mixing interaction occurs if 
there are small mirror matter bodies in our solar system.
It has been suggested in Ref.\cite{tunguska,puz,yoon} that the 
collisions of such bodies with the Earth may naturally explain
many puzzling features of the 1908 Tunguska event as well as 
the low altitude anomalous fireball events.
An important implication of this explanation is that mirror
matter fragments should exist in the ground at these
impact sites. Clearly, an important issue is how to search
for mirror matter in the ground. One concrete idea was
proposed in Ref.\cite{yoon}, where it was suggested that
mirror matter in the ground can cool the surrounding ordinary
matter by drawing in heat and radiating it away as mirror
photons\footnote{
Actually, the cooling effect of mirror matter that
is vapourized in the atmosphere (from e.g. small
mirror matter space-bodies or dust particles colliding 
with the Earth) might also lead to
interesting effects. For example, one might think that
rapid cooling of the atmosphere due to mirror matter
absorbing heat from the surrounding ordinary atoms (and
radiating that heat way as mirror photons) might 
lead to the formation of clouds and maybe even ice blocks. 
Perhaps this might be connected with long
standing observations 
of atmospheric holes\cite{smallcomets} and observations
of falling ice blocks\cite{spain}.}.
In this paper we will discuss a complementary way
of searching for mirror matter in the ground. 

While the force of gravity can be opposed by electrostatic
forces for a solid piece of mirror matter (if $|\epsilon| \sim 10^{-6}$), 
one might expect
mirror matter in gaseous form to eventually diffuse to the center
of the Earth. This is only true, though, if the mirror atoms 
cannot bind with ordinary atoms at ordinary temperatures.
If the effective ordinary electric charge of the mirror nuclei is of
opposite sign to ordinary nuclei (i.e. $\epsilon < 0$), 
then ordinary - mirror
nuclei can form bound states. Consider the case of a 
proton and a mirror proton. In this case, the Bohr radius
of the proton - mirror proton bound state ($R_b$) and its
binding energy ($E_b$) are given by (we use natural
units where $\hbar = c = 1$ throughout):
\begin{eqnarray}
R_b &=& {m_e \over \mu \epsilon}r^H_{bohr} = 
{ 1 \over \mu |\epsilon| e^2} \sim 10^3\left( {10^{-6} \over |\epsilon|}
\right)r^H_{bohr}, \nonumber \\
E_b &=& {\epsilon^2 \mu \over m_e}E^H_{bohr} = {-\mu \epsilon^2 e^4 \over 2}
\sim -10^{-8}\left( {\epsilon \over 10^{-6}}\right)^2 \ eV,
\end{eqnarray}
where $\mu = m_p/2$ is the reduced mass and $r^H_{bohr}, E^{H}_{bohr}$
are the standard Bohr radius and ground state energy for Hydrogen.
Clearly, the binding energy is too low to be of
much interest if $|\epsilon| \approx 10^{-6}$ as
suggested by the orthopositronium experiments. However, 
the binding energy increases
for larger masses and also for larger charges.

Perhaps an interesting observation is that if the Bohr radius of the
ordinary - mirror nuclei bound state is less than the Bohr radius
of the inner electrons and mirror electrons, then the electric charge
of the ordinary nuclei and mirror nuclei are, 
to a good approximation, {\it not} screened by
the electrons and mirror electrons. 
This means that we can, to a good approximation,
ignore the electrons and mirror electrons and treat atoms as nuclei. 
The condition for this to occur
is that:
\begin{eqnarray}
& R_b  \stackrel{<}{\sim} r^K_b
& \Rightarrow
{1 \over \mu |\epsilon| Z_1 Z_2 e^2} \stackrel{<}{\sim}
{1 \over m_e e^2 Z_1}
\nonumber \\
&{\rm That \ is, \ }&
Z_2 \stackrel{>}{\sim} {m_e \over \mu}{1 \over |\epsilon|} ,
\end{eqnarray}
where we have assumed $Z_1 \ge Z_2$ and used the notation $r^K_b$ for
the Bohr radius of the inner K-shell electrons. 
For light elements, $Z_1, Z_2 < 15$, the above condition is never
satisfied; which means that ordinary - mirror atom 
light element bound states do not
exist or have extremely low binding energies (certainly too
low to exist on Earth). However, for
heavy elements, $Z_1, Z_2 \stackrel{>}{\sim} 15$, the above
condition is satisfied.
Taking as an example, $Fe - Fe'$ bound state, then 
$R_b = 0.7 \ r^K_b$, verifying that the ordinary and mirror
atom are bound so closely that the screening effects of the ordinary
and mirror electrons can be approximately ignored.
The binding energy of the
ground state in the $Fe-Fe'$ system is approximately:
\begin{eqnarray}
E_{Fe-Fe'} &\simeq & {-\epsilon^2 m_{Fe} Z_{Fe}^4 e^4 \over 4}
\nonumber \\
&\simeq & 0.3 \left( {\epsilon \over 10^{-6}}\right)^2 \ eV .
\end{eqnarray}
While this particular bound state could have interesting
implications in cold environments such as the ISM,
it would not exist indefinitely at ordinary temperatures for 
single $Fe'$ atoms.
%\footnote{ It may however be important in the case of a mirror iron 
%fragment (or mirror iron microparticle) which happened to be 
%stopped in ordinary iron. In that case, some fraction of the iron atoms 
%may reside in the ground state, however interactions
%between the mirror iron atoms would need to be considered as well.}.
They would occasionally be excited by thermal interactions, eventually
diffusing towards the Earth's center. However,
heavier bound states have higher binding energies and
can be quite stable at ordinary temperatures.
For example, the binding energy of $Pb - Fe'$ is about
5 eV and $Pb - Pb'$ is about 120 eV.

The above observations open several new ways to search
for mirror matter. For example, if mirror iron meteorites
exist, then such bodies can vaporise in the atmosphere.
Each iron atom would 
diffuse towards the ground, where it would only be stopped 
(permanently) if
it encountered an ordinary element heavier than $Fe$, e.g. lead
(or it encounted some embedded mirror matter fragment). 
Thus, ordinary samples of lead, may contain a small fraction
of $Pb-Fe'$ bound states which have built up over time.
[Mirror iron may also come from cosmic rays, since we know
that ordinary cosmic rays contain $Fe$, it is very plausible that
they might also contain $Fe'$.]

The capture cross section of say, $Fe'$ on lead can
be estimated to be \footnote{
This cross section can be obtained from the standard 
formula for the capture of a charge electron on
a nucleus (see e.g. Ref.\cite{standard}), after the 
replacements $e \to Z_{Fe} e$,\ $Z_{\mbox{nucleus}}\to \epsilon Z_{Pb}
$ and $m_e \to m_{Fe}$.}:
\begin{eqnarray}
\sigma_{rc} = {256 \pi^2 \over 3 m_{Fe}^{2}} Z_{Fe}^2 
\alpha {\xi^6 \over (1 + \xi^2)^4}
{exp(2\xi \tau) \over 1 - exp(-2\pi\xi)} 
\end{eqnarray}
where $\sin \tau = {-2\xi\over 1 + \xi^2}$ and 
$\xi = \epsilon \alpha Z_{Fe} Z_{Pb}/v$.
In order to work out $\sigma_{rc}$ for the capture of $Fe'$ on lead,
we need to first estimate the velocity, $v$, of the $Fe'$ atoms.
For the interesting case of $|\epsilon| \approx 10^{-6}$, the
$Fe'$ undergoes many elastic collisions with the ordinary
nuclei so that by the time it
could reach any lead deposit in the ground, an individual $Fe'$ atom
would have thermalized with the surroundings. 
This means that the mean energy of an individual $Fe'$
atom would be approximately, 
\begin{eqnarray}
&\langle E \rangle & \approx \frac{3}{2}kT, \nonumber \\
&\Rightarrow &
v/c \approx 
\sqrt{ 3kT \over m_{Fe}} \approx 10^{-6}.
\end{eqnarray}
If $|\epsilon| \approx 10^{-6}$ as
suggested by the orthopositronium experiments then
$\xi \approx 13$ and $\sigma_{rc}$ is
\begin{eqnarray}
\sigma_{rc} \approx 3\times 10^{-30} \ cm^2
\end{eqnarray}
This means that $Fe'$ must travel an average total length of
$\ell = 1/n_{Lead}\sigma_{rc}
\approx
3\times 10^{7} km$ before being captured by lead
(where we have used the mean number density of lead atoms in the
Earth's crust of $n_{lead} \sim 10^{17}/cm^3$). 
 The average time needed for $Fe'$
to be captured by lead is thus about $\ell/v\approx 3$ years. 
Because the elastic scattering with nuclei is much larger
($\sigma \sim 10^{-20}\ cm^2$), $Fe'$ will be captured by lead at a 
distance much smaller than $\ell$.
An $Fe'$ atom has a mean free path in the earth's crust
of about $\ell_{m}=1/n\sigma\sim 10^{-3} cm$. The $Fe'$ atom
will thus typically perform a random walk of about 
$N=\ell/\ell_{m}\sim 3\times 10^{15}$ steps of length
$\ell_{m}$. The typical distance a $Fe'$ atom will stray 
before being captured by
lead is thus about $\sqrt{N}\times \ell_{m}\approx 500$ meters.

It is also possible for $Fe'$ (and other potential heavy mirror
elements) to be absorbed by colliding with any mirror matter fragments
that happen to be embedded in the Earth.
The strength of the capture process relative to condensation on mirror
matter fragments is of course difficult to quantify. For this reason
and because the (time) integrated 
flux of $Fe'$ atoms striking the Earth (from e.g. the 
vapourization of $Fe'$ meteorites in the atmosphere or
from cosmic rays if they contain an $Fe'$ component) is essentially
unknown, it is impossible to give precise predictions for the 
exotic ordinary - mirror atom
bound state abundances in materials containing heavy elements
such as lead.

Interestingly, a recent experiment has searched for bound states 
consisting of
strongly interacting massive particles (SIMPS) and iron and 
gold nuclei \cite{jav}.
No such bound states were found. Using these results an 
upper limit of about $10^{-11}$
can be set for $Au-Fe'$ bound states abundances, and an upper 
limit of $10^{-10}$ for $Au-Pb'$
bound states abundances. Samples near the Earth's surface 
were tested, because SIMPS (being strongly interacting) are 
assumed to have a
penetration depth of a few meters or less. Since a mirror atom is expected to penetrate
much deeper into the Earth before being captured, it is not
necessary to focus on samples near the Earth's surface.
Any sample containing heavy elements such as lead could
be used. In particular, a much more senstive search for
ordinary-mirror atom bound states could be preformed
by working with
liquid compounds such as lead tetra chloride ($Pb Cl_4$). Such liquids
could be put into a centrifuge, thereby
greatly enhancing the concentration of, say, $Pb-Fe'-Cl_4$
(in this example). After this purification procedure the
sample could be then put in a mass spectrometer. In this
way a very sensitive limit on the abundance of such exotic
states could be achieved (perhaps a sensitivity of order
$10^{-16}$ could be expected).

While our discussion has focused on the capture of
single $Fe'$ atoms, interesting effects can also occur for
$Fe'$ fragments (and other heavy mirror elements), including 
tiny micron-sized mirror iron
particles containing e.g. $\sim 10^8 \ Fe'$ atoms.
Such fragments would be stopped in the Earth after only a very
short distance (typically less than a millimeter). 
Thus, unlike the case of individual atoms, small fragments
and microparticles might be expected to occur only 
in lead samples exposed to the atmosphere.
The binding energy of the fragment depends on several
factors including its crystal structure and lattice
spacing. At ordinary temperatures lead forms a face-centered cubic
structure while iron forms a body-centered cubic structure.
Depending on how these lattices are arranged relative
to each other, there may be several possible stable configurations.
The state that an embedded fragment happens to be in might
not be the lowest energy configuration. This point is
illustrated in Figure 1, where the system is in the second 
local minimum. 
In this case it might be possible to cause
a phase transition to a lower energy configuration by irradiating
the sample with a laser resulting in emission of UV
photons with higher energy than the laser photons
if the final configuration involves $Pb-Fe'$ bound states.
Interestingly, there is a recent experiment which has indeed
found such a puzzling burst of UV photons from laser irradiated
lead targets\cite{laser}. Further work needs to be done,
both experimentally and theoretically to see if this
mirror matter explanation could really be the origin of
those observations.

In addition to terrestrial implications, one might also 
imagine interesting astrophysical implications of ordinary-mirror
mixed matter.
If, for example, the interstellar medium (ISM) has dust particles 
containing both ordinary and mirror fragments.
The interstellar medium is a very cold environment which means
that even relatively light states such as $Fe - Fe'$ or $Fe-Mg'$ 
could exist leading
to infra-red emission lines. Also heavier states such as $Pb-Fe'$ (or
$Pb'-Fe$) could
also exist but would have absorption in the UV. Interestingly there are 
many unidentified infra-red, optical and UV lines in the ISM (for
example the well known UV absorption feature at $\lambda = 2175 \ A$), but we
postpone a more detailed analysis for the future. 

In conclusion, we have pointed out that the small photon - mirror
photon kinetic mixing interaction -- suggested by orthopositronium
experiments --
should cause single heavy mirror atoms (such as $Fe'$) to become 
bound with heavy
ordinary atoms. These bound states could be found if
sensitive mass spectroscopy was done on materials such as ordinary lead.
The existence of these bound states could also lead to
interesting
implications if small mirror fragments (such as iron microparticles)
are stopped in ordinary materials containing heavy elements such as lead.
We have speculated that this might explain a recent laser
experiment which has found an unexpected burst of UV light from
laser irradiated lead targets. The ordinary - mirror atom
bound states may also have interesting astrophysical implications 
if such mixed matter exists in the ISM.

\vskip 1.2cm \noindent {\bf Acknowledgements} \vskip 0.5cm 
We would like to thank A. Pakhomov for useful correspondence
regarding his experiment. R.F. wishes to acknowledge very useful
and interesting discussions/correspondence with 
J. Learned, J. Martinez-Frias,
S. Pakvasa, R. Volkas, T. L. Yoon and wishes to thank
T. Mesirow for bringing the laser experiment to his attention.

\newpage
{\Large \bf Figure Caption}
\vskip 1cm

\noindent
{\bf Figure 1:} 
Possible form of the potential energy of a mirror
iron fragment embedded in a piece of lead.

\newpage
\begin{center}
\includegraphics[width=0.5\textwidth]{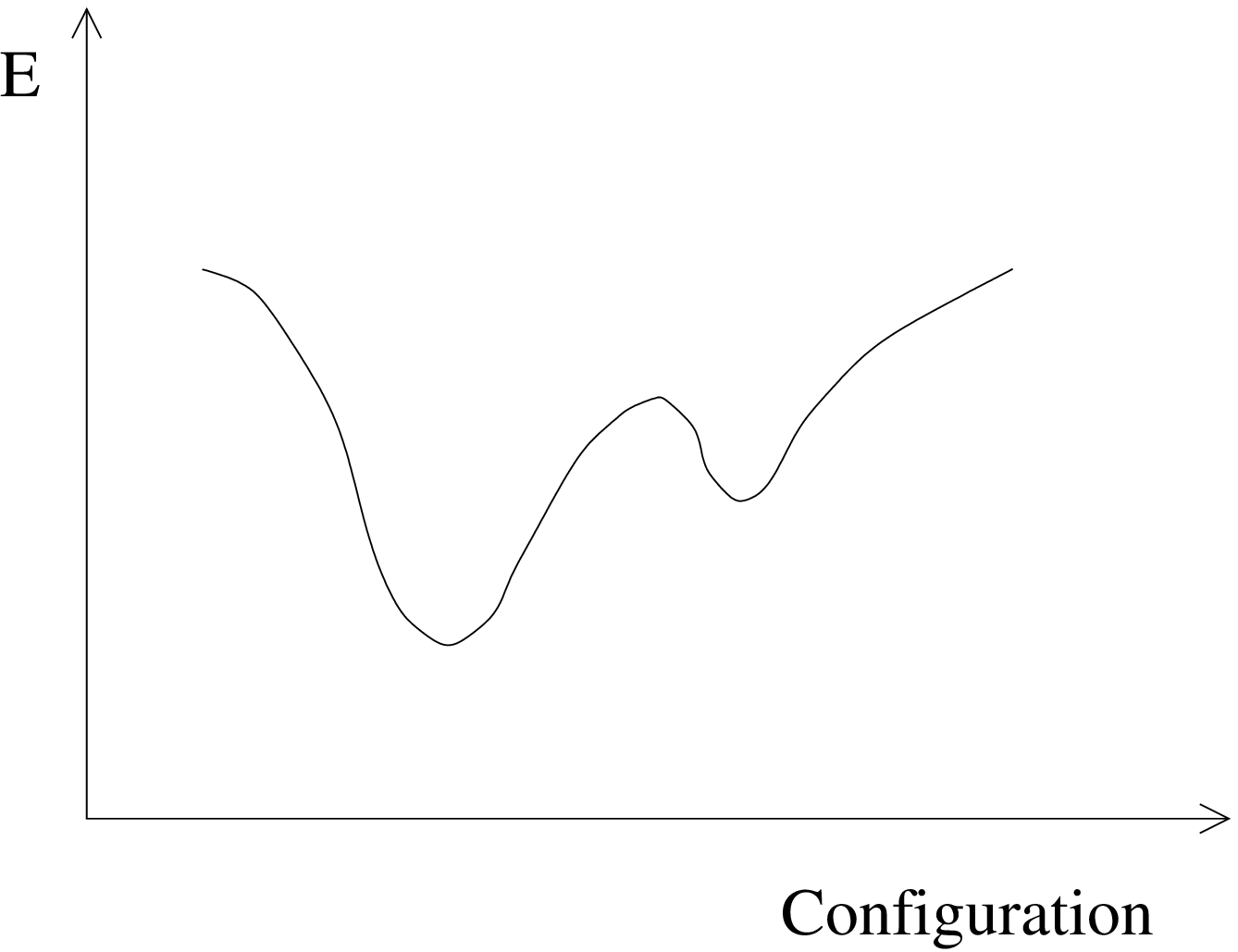}
\end{center}
\end{document}